\documentclass[prd]{revtex4}
\usepackage{graphicx}

\begin{document}

\title{Finite Temperature Induced Fermion Number\\ In 
The Nonlinear $\sigma $ Model In (2+1) Dimensions }

\author{Gerald V. Dunne$^a$, Justo Lopez-Sarrion$^b$ and Kumar Rao$^a$}

\affiliation{$^a$Department of Physics, University of Connecticut, 
Storrs CT
06269-3046, USA\\
$^b$Departamento de Fisica, Universidad de Zaragoza,
E-50009 Zaragoza, Spain}


\begin{abstract}
We compute the finite temperature induced fermion number for fermions coupled to
a static nonlinear sigma model background in $(2+1)$ dimensions, in the derivative
expansion limit. While the zero temperature induced fermion number is well
known to be topological (it is the winding number of the background), at finite
temperature there is a temperature dependent correction that is nontopological --
this finite $T$ correction is sensitive to the detailed shape of the
background. At low temperature we resum the derivative expansion to all orders, and
we consider explicit forms of the background as a $\mathcal{CP}^1$ instanton or as
a baby skyrmion.

\end{abstract}

\maketitle

\section{Introduction}

The phenomenon of induced fermion number arises due to the interaction of
fermions with nontrivial topological backgrounds (e.g., solitons, vortices,
monopoles, skyrmions), and has many applications ranging from polymer physics
to particle physics \cite{jr,gw,bag,jackiw2,rajaraman,niemi,polymer,amo}.
At zero temperature, the induced fermion number is a topological quantity
(modulo spectral flow effects), and is related to the spectral asymmetry
of the relevant Dirac operator, which counts the difference between the
number of positive and negative energy states in the fermion spectrum.
Mathematical results, such as index theorems and Levinson's theorem,
imply that the fractional part of the zero temperature induced fermion 
number is topological in the sense that it is determined by the
asymptotic properties of the background fields
\cite{rich,wilczek,boy,mike,poly,jaffe2}. This topological character of the induced
fermion number is a key feature of its application in certain model field theories
\cite{eric,weigel,manu,diakonov}. At finite  temperature, the situation is very different -- the
induced fermion number is  generically nontopological, and is not a sharp observable
\cite{ad,dr,leipzig}. Several explicit examples of kinks \cite{ns,soni,keil} and sigma
models \cite{ad,dr} in $(1+1)$ dimensions, and monopoles \cite{cp,goldhaber,monopole,ad}
in $(3+1)$ dimensions, have been analyzed in detail. In this paper we compute the
finite  temperature induced fermion number for fermions coupled to a static nonlinear
sigma model background in $(2+1)$ dimensions. 

At zero temperature, the induced fermion number for fermions coupled
to a static nonlinear sigma  model background in $(2+1)$ dimensions has
been studied extensively
\cite{jaro1,carena,aw,abanov}. The $T=0$ induced fermion number is equal to the
winding number of the sigma model background field, and this result may be
interpreted in terms of a topological current density. This system has applications
both in condensed matter physics \cite{wilczekzee,aw}, where the sigma model
provides a phenomenological model for a two dimensional Heisenberg ferromagnet, and
in particle and nuclear physics, where the sigma model can be used as a ``baby
Skyrmion'' model \cite{jaro1,carena,jaro2,krishna} that mimics many
properties of the $(3+1)$ dimensional Skyrme model for baryons
\cite{weigel}. One motivation for this paper
is to compute the induced fermion number at nozero temperature and to
understand in detail the origin of the nontopological $T$ dependent
contributions, as similar effects will occur in the $(3+1)$ dimensional
Skyrme case.

In Section II we define what is meant by finite temperature induced fermion
number, and indicate how it can be computed. In Section III we use the derivative
expansion to compute the finite temperature induced fermion number for a nonlinear
$\sigma$-model background in $(2+1)$ dimensions. At low temperature we resum the
derivative  expansion to all orders, and we consider explicit forms of the background
as a $\mathcal{CP}^1$ instanton or as a baby skyrmion. We conclude in Section IV
with some comments concerning the relation of our results to the well
known $T=0$ results, and concerning the possible extension to $(3+1)$
dimensional Skyrme models.

\section{Finite Temperature Induced Fermion Number}

The induced fermion number is an expectation value of the second
quantized fermion operator $N=\frac{1}{2}\int
dx\,[\Psi^\dagger,\Psi]$.  For a given static classical background
field configuration, the fermion field operator $\Psi$ can be expanded
in a complete set of eigenstates of the Dirac Hamiltonian $H$. 
Expectation values of $N$ can then be computed. At zero temperature,
the fermion number is a vacuum expectation value
$\langle N\rangle_0\equiv\langle 0\vert N\vert 0
\rangle$, and is related to the spectral asymmetry of the Dirac
Hamiltonian \cite{niemi}
\begin{eqnarray}
\langle N\rangle_0&=&-\frac{1}{2}\,(\textrm{spectral asymmetry})\nonumber\\
&=&-\frac{1}{2}\int_{-\infty}^{\infty} dE \, \sigma(E)
\,\,\textrm{sign}(E)
\label{asymmetry}
\end{eqnarray}
Here $ \sigma (E)$ is the spectral function of the Dirac Hamiltonian
$H$ :
\begin{equation}
\sigma (E)=\frac{1}{\pi}\textrm{ Im Tr}
\,\left(\frac{1}{H-E-i\epsilon}\right)
\label{spectral}
\end{equation}
At nonzero temperature, $T$, the induced fermion number is a
{\it thermal} expectation value :
\begin{eqnarray}
\langle N\rangle_T &=&\frac{\textrm{Tr}\,(e^{-\beta
H}N)}{\textrm{Tr}\,(e^{-\beta H})}\nonumber\\
                   &=&-\frac{1}{2}\int_{-\infty}^{\infty}dE \,\sigma (E)
                   \,\tanh(\frac{\beta  E}{2})
\label{nt}
\end{eqnarray}
where $\beta \equiv \frac{1}{T}$. Notice that this finite temperature
expression (\ref{nt}) reduces smoothly to the zero temperature expression
(1) as $\beta\to \infty$. In fact, the nonzero temperature provides a
physically meaningful and mathematically elagant regularization of the
spectral asymmetry. 

The spectrum of the fermions is independent of the temperature in our approximation of
a static classical background field. All information about the fermion spectrum, in
the presence of the background, is encoded in the spectral function 
$\sigma(E)$ defined in
(\ref{spectral}). Thus, knowledge about the spectral function
$\sigma(E)$ is the key to evaluating (\ref{asymmetry}) or
(\ref{nt}). Actually, to compute the induced fermion number (either at
$T=0$, or $T>0$) one only needs the {\it odd} part of the spectral
function, as is clear from (\ref{asymmetry}) and (\ref{nt}). Thus, the
calculational problem is to find the odd part of $\sigma(E)$.

A more physical interpretation of the finite $T$ fermion number
$\langle N\rangle _T$ is provided by noting \cite{dr,leipzig} that 
$\langle N\rangle_T$ in (\ref{nt}) separates naturally into a
$T=0$ piece and a finite $T$ correction
\begin{equation}
\langle N\rangle _T=\langle N\rangle _0 +\int_{-\infty}^{\infty}dE \,\sigma (E) 
\,\textrm{sign}(E)\, n\,(|E|)\quad ,
\label{split}
\end{equation}
where
\begin{equation}
n(E)=\frac{1}{e^{\beta E}+1}
\label{fd}
\end{equation}
is the Fermi-Dirac distribution function, and we have used the simple fact that
$\tanh(\frac{\beta E}{2})=1-2n(E)$.

The first term, $\langle N \rangle_0 $, in (\ref{split}) is known to be topological 
in the sense that it is invariant under small local deformations of the background
which do not change the asymptotic boundary values of the background. Physically,
this topological term corresponds to the vacuum polarization response of the system
to the presence of the background. The second term in (\ref{split}), which is the
finite $T$ correction term, is generically nontopological in the sense that it is
sensitive to local deformations in the background. Physically, this term corresponds
to the plasma response of the system to the presence of the background. It
incorporates effects beyond the vacuum response, and so is more sensitive to the
entire single-particle fermion spectrum. The caveat ``generically'' has been included
here because for certain very special backgrounds the finite $T$ correction is itself
topological. This happens when the Dirac Hamiltonian has a quantum 
mechanical supersymmetry
relating the positive and negative parts of the spectrum \cite{dr,leipzig}. Several
explicit examples (in $(1+1)$ and $(3+1)$ dimensions) of this distinction
between topological and nontopological contributions have been studied
in detail already \cite{ad,dr,leipzig}. 
An example in $(2+1)$ dimensions where the finite $T$ correction is
topological is the case of a background flux string
\cite{cesar,cp,monopole}. In this paper we study a
$(2+1)$ dimensional example in which the finite $T$ correction is
nontopological. 

\section{The Non Linear Sigma Model In (2+1) Dimensions}

Consider fermions coupled to a static $\sigma$-model background in $(2+1)$ dimensions.
\begin{equation}
{\cal L}=\bar{\psi}(i\gamma^{\mu}\partial_{\mu}-m\hat{n}\cdot \vec{\tau})\psi
\label{lag}
\end{equation}
The fermions are in the defining representation of SU(2), and the static
$\sigma$-model background is represented by a 3-vector (in internal
space) $\hat{n}(\vec{ x})$, which is constrained to take values on $ S^{2}$:
\begin{equation}
\hat{n}^{2}(\vec{x})=1
\label{constraint}
\end{equation}
In the Lagrangian (\ref{lag}), the $\tau^{a}$ are Pauli matrices,
and $m$ is a mass parameter that sets an important energy scale of the
single-particle fermion spectrum. Later in the paper we will consider
specific profiles for
$\hat{n}(\vec{x})$.  At
$T=0$ the quantization of fermions in this background is well known,
and leads to an induced fermion number
\cite{jaro1,carena,aw}
\begin{equation}
\langle N\rangle_0=\frac{1}{8\pi}\,  \int d^{2}x\,  \epsilon^{abc}\, 
\epsilon^{ij}\, \hat{n}^{a}\,\partial_i\hat{n}^{b}\, \partial_j\hat{n}^{c}
\label{nzero}
\end{equation}
This induced fermion number, $\langle N\rangle _0$, is just the integer
winding number of the map $\hat{n}: \Re^{2}\to S^2$. With the boundary
condition that $\hat{n}(\vec{x})$ has an angle independent limit at
spatial infinity, we can compactify the spatial $\Re^{2}$ to $S^{2}$,
in which case $\langle N\rangle _0$ gives the integer winding number
for maps $\hat{n}:S^{2}\to S^{2}$. Thus, $\langle N\rangle _0$ is
topological.

Invoking Lorentz invariance, one interprets the result (\ref{nzero}) as the spatial
integral, $\int d^{2}x J^{0}$, of the zeroth component $J^{0}$ of a ``topological''
current density
\begin{equation}
J^{\mu}=\frac{1}{8\pi}\,\epsilon^{\mu\nu\lambda}\, \epsilon^{abc}\, \hat{n}^{a}\,  
\partial_{\nu}\hat{n}^{b}\, \partial_{\lambda}\hat{n}^{c}
\label{top}
\end{equation}
This current is manifestly conserved ($\partial_{\mu}J^{\mu}=0$) without use of the
equations of motion, and hence is called ``topological''. In fact, $J^{\mu}$ may (and
generally does) have higher derivative corrections beyond the term shown in
(\ref{top}). However, these are all total derivatives and hence do not contribute to
an integrated quantity such as the zero temperature fermion number $\langle
N\rangle_0$ in (\ref{nzero}) \cite{wilczek}. 

At nonzero T, several aspects of this familiar story change. First, as
we show explicitly below for this $(2+1)$ dimensional $\sigma$-model
case, there are higher derivative corrections to $J^0$ that are not
total derivatives, and these contribute to the finite $T$ correction
to the $T=0$ induced fermion number (\ref{nzero}). Second, at finite T
one can no longer invoke Lorentz invariance (the thermal bath
determines a preferred frame), and so one cannot directly infer a
topological current density as was done in (\ref{top}). 

A convenient approach to this calculation is to express (\ref{nt}) as a
contour integral (using the $ i\epsilon$ prescription in the spectral
function) in terms of the Dirac resolvent
\begin{equation}
R(z)\equiv \textrm{ Tr}(\frac{1}{H-z})
\label{resolvent}
\end{equation}
Thus we have
\begin{equation}
\langle N\rangle _T =-\frac{1}{2}\int_\mathcal{C}\frac{dz}{2\pi i}\,
\textrm{Tr}\, (\frac{1}{H-z}) \,\tanh(\frac{\beta z}{2})
\label{contour}
\end{equation}
where $\mathcal{C}$ is the contour ($-\infty +i\epsilon ,+\infty +i\epsilon$) and 
($+\infty -i\epsilon ,-\infty -i\epsilon$) in the complex energy plane. For a
hermitean Hamiltonian $H$, the poles and branch cuts of the resolvent lie on the
real axis, so one has two choices for evaluating the contour integral in
(\ref{contour}). First, one can deform the contour $\mathcal{C}$ around the poles and
cuts of the resolvent $R(z)$. This approach leads to an integral representation for
$\langle N\rangle_{T}$. Alternatively, one could deform the contour $\mathcal{C}$
around the Matsubara poles, $z_n=(2 n+1)\,i\pi  T$, of the $\tanh(\frac{\beta
z}{2})$ function, which lie on the imaginary axis. This leads to a summation
representation for $\langle N\rangle _T$. These two representations are, of course,
equivalent. The integral representation is just the familiar Sommerfeld-Watson
transform of the summation representation.  

Thus, in order to evaluate the finite temperature induced fermion
number,
$\langle N\rangle_T$, one needs the odd part of the spectral function
$\sigma(E)$; or, equivalently, the even part of the resolvent $ R(z)$. No
exact expression is known for 
$R_{\rm even}(z)$ for the interaction described by (\ref{lag}).
Therefore, one needs an approximate technique. Here we use the
derivative expansion \cite{ian}, in which one assumes that the natural
length scale $\lambda$ associated with the background field,
$\hat{n}(\vec{x})$, is large compared to the Compton wavelength
$\frac{1}{m}$ of the fermion field: $\lambda m \gg 1$. That is, we
assume that
\begin{equation}
|\vec{\nabla}\hat{n}(\vec{x})|\ll m
\label{deriv}
\end{equation}
By symmetry, this means that we assume that all derivatives of $\hat{n}$ are small
compared to the fermion mass scale $m$. Since the background is static,
these derivatives are all spatial derivatives. The derivative expansion
of the resolvent (and hence of the induced fermion number) can then be
formalized as follows. The Dirac Hamiltonian corresponding to the
Lagrangian (\ref{lag}) is
\begin{equation}
H=-i\vec{\alpha}\cdot \vec{\nabla}+m\, \hat{n}\cdot\vec{\tau}\,\gamma^{0}
\label{ham}
\end{equation}
Here, $(\vec{\alpha}=\gamma^{0}\vec{\gamma})$, and for definiteness we choose gamma
matrices as follows: $\gamma^{0}=\sigma^{3},\,\,
\vec{\gamma}=-i\vec{\sigma}$. The even part (in $z$) of the resolvent (\ref{resolvent})
is
\begin{equation}
\left[\textrm{Tr}\left(\frac{1}{H-z}\right)\right]_{\rm
even}=\textrm{Tr}\left(H\,\frac{1}{H^{2}-z^{2}}\right)
\label{evenres}
\end{equation}
This can be systematically expanded in terms of derivatives of the background field
as follows. First, write the square of the Hamiltonian (\ref{ham}) as
\begin{eqnarray}
H^2&=&-\vec{\nabla}^{2}+m^{2}+im\gamma^{j} \partial_j\hat{n}\cdot
\vec{\tau}\nonumber\\
&\equiv& H_0^{2}+V
\end{eqnarray}
where $H_0^{2}=-\vec{\nabla}^{2}+m^{2}$ is the square of the free
Hamiltonian, and the ``interaction'' is $V=im\gamma^{j}
\partial_j\hat{n}\cdot
\vec{\tau}$. Next, since $V$ involves a factor of $m$ and a derivative
of $\hat{n}$, by (\ref{deriv}) it is small relative to the natural
scale, $m^2$, of $H_0^2$. Therefore, the even part of the resolvent in
(\ref{evenres}) can be systematically expanded in terms of
$V$ : 
\begin{eqnarray}
\left[\textrm{Tr}\left(\frac{1}{H-z}\right)\right]_{\rm{even}}=\textrm{Tr}(\triangle
V\triangle V\triangle I)-
\textrm{Tr}(\triangle V\triangle V\triangle V\triangle K)+\textrm{Tr}(\triangle
V\triangle V\triangle V\triangle V\triangle I)- \cdots 
\label{expansion}
\end{eqnarray}
where we have defined the kinetic part of $H$ as
$K=-i\vec{\alpha}\cdot \vec{\nabla}$, the interaction part of $H$ as 
$I=m\hat{n}\cdot
\vec{\tau}\gamma^{0}$, and the free propagator as
\begin{equation}
\triangle\equiv \frac{1}{-\vec{\nabla}^{2}+m^{2}-z^{2}}
\label{freeres}
\end{equation}
In obtaining (\ref{expansion}) we have used the simple facts
that $\textrm{Tr}(\tau^{a})=0$ and $\textrm{Tr} (\gamma^{i_1}\dots
\gamma^{i_{2n+1}}\gamma^{0})=0$. The trace in (\ref{expansion})
involves a matrix trace over the Dirac matrices, $\gamma
^{\mu}=(\sigma^{3}, -i\vec{\sigma})$, and  the isospin matrices,
$\vec{\tau}$, as well as a functional trace over the propagators
$\triangle$. These traces can be done in momentum space or in position
space. In position space, as we go to higher derivative orders we make use of the
following operator identity:
\begin{eqnarray}
\frac{1}{-\vec{\nabla}^{2}+a^{2}} V(\vec{x})&=&V(\vec{x})\frac{1}{-\vec{\nabla}^{2}+
a^{2}} +2\left(\vec{\nabla}V(\vec{x})\right)\cdot
\vec{\nabla}\frac{1}{(-\vec{\nabla}^{2}+a^{2})^{2}}+\left(\vec{\nabla}^{2}
V(\vec{x})\right)\frac{1}{(-\vec{\nabla}^{2}+a^{2})^{2}}                                                                                                    
\nonumber\\ & & + 4\left(\partial_{i} \partial_{j}  V(\vec{x})\right)\partial_{i} \partial_{j}
\frac{1}{(-\vec{\nabla}^{2}+a^{2})^{3}}+4\left[\vec{\nabla}\left(\vec{\nabla}^{2} V(\vec{x})\right)\right] \cdot
\vec{\nabla} \frac{1}{(-\vec{\nabla}^{2}+a^{2})^{3}}+\cdots
\label{spatial}
\end{eqnarray}
where the parentheses around the derivatives on the RHS of
(\ref{spatial}) indicate that the gradient operator acts on
$V(\vec{x})$ only. At any given order of the derivative expansion we
collect together the required number of derivatives of
$\hat{n}(\vec{x})$ , as given by repeated application of the above
identity, noting that
$V(\vec{x})$ itself involves one derivative of $\hat{n}(\vec{x})$.
This,  however,  becomes rather clumsy for higher order terms in the
derivative expansion and in practice it is easier to do the
computation in momentum space.    The leading
term in the derivative expansion  involves two
derivatives, which come directly from the two factors of $V$ in the
first term of (\ref{expansion}) :
\begin{eqnarray}
\label{fn2}
\left[\textrm{Tr}\left(\frac{1}{H-z}\right)\right]_{\rm{even}}^{(2)}&=&\textrm{Tr}\left(\triangle V\triangle 
V\triangle I\right)^{(2)}\nonumber \\
&=&\frac{-m^{3}}{2 \pi(m^{2}-z^{2})^{2}} \int d^{2}x \, 
\epsilon^{ij}\, \epsilon^{abc}\, \hat{n}^{a}\, \partial_i\hat{n}^{b}\, \partial_j\hat{n}^{c}
\end{eqnarray}
To compute the induced fermion number to this (leading) order of the
derivative expansion, we evaluate the $z$ integral in
(\ref{contour}) as a sum over Matsubara modes. This leads to the
following contribution to the induced fermion  number:
\begin{eqnarray}
\label{derivative2}
\langle N\rangle _T^{(2)} & = &   \left(\sum_{n=-\infty}^\infty
\frac{m^3 T}{2\pi  [m^2 +((2 n+1)\pi T)^2]^2}\right) \, \int d^{2}x\, 
\epsilon^{ij}\, 
\epsilon^{abc}\, 
\hat{n}^{a}\, \partial_i \hat{n}^{b}\, \partial_j \hat{n}^{c}  
\nonumber \\  {}  
& = & 
\left(\sum_{n=-\infty}^\infty \frac{4 m^3 T}{[m^2 +((2n+1)\pi T)^2]^2}
\right)\,\,
\langle N\rangle_0 
\label{nt2}
\end{eqnarray}
The prefactor in (\ref{nt2}) is temperature dependent, but in
the 
$T\rightarrow 0$ limit it reduces smoothly to one:
\begin{equation}
\sum_{n=-\infty}^\infty \frac{4 m^3 T}{[m^2 +
((2n+1)\pi T)^2]^2}=1-\left(\frac{2 m}{T}\right)e^{-\frac{m}{T}}+\dots
\label{2der}
\end{equation}
Here we have used the general low temperature expansion:
\begin{equation}
T\sum_{n=-\infty}^\infty \frac{1}{(((2n+1)\pi T)^{2}+m^{2})^{p}} \sim 
\frac{m^{1-2p}}{2\sqrt{\pi}} \frac{\Gamma (p-\frac{1}{2})}{\Gamma
(p)}-\frac{(2mT)^{1-p}}{m\Gamma (p)} e^{-\frac{m}{T}}+ \cdots 
\label{identity}
\end{equation}
Thus, in the  $T\to 0$ limit, the leading derivative expansion term
(\ref{nt2}) produces the entire zero temperature answer, $\langle
N\rangle_0$. And at nonzero temperature, this order of the derivative
expansion gives the result (\ref{nt2}), which is topological because
it is simply the $T=0$ result multiplied by a smooth function of
temperature.

To see the nontopological contribution to $\langle N\rangle_T$ we need
to go to the next order in the derivative expansion, which involves
four derivatives of $\hat{n}(\vec{x})$. Recalling that
$V=im\gamma^{j}\partial_{j}\hat{n}(\vec{x})\cdot\vec{\tau}$ already
includes one derivative of $\hat{n}(\vec{x})$, we see that at fourth
order we can get contributions from the first three terms in
(\ref{expansion}):
\begin{equation}
\left[\textrm{Tr}\left(\frac{1}{H-z}\right)\right]_{\rm{even}}^{(4)}=\textrm{Tr}(\triangle
V \triangle  V \triangle I)^{(4)}-\textrm{Tr}(\triangle V \triangle V
\triangle V \triangle K)^{(4)} +\textrm{Tr}(\triangle V \triangle V
\triangle V \triangle V \triangle I)^{(4)} 
\label{expansion4}
\end{equation}
After a straightforward expansion one finds
\begin{eqnarray}
\left[\textrm{Tr}\left(\frac{1}{H-z}\right)\right]_{\rm{even}}^{(4)} &
= & -\frac{m^{3}}{12 \pi} \frac{m^{2}+ 5z^{2}}{(m^{2}-z^{2})^{4}} \int
d^{2}x\,(\epsilon^{ijk}\, \epsilon^{abc}\,
\hat{n}^{a}\,\partial_i\hat{n}^{b}\,\partial_j\hat{n}^{c})(\partial_k\hat{n}^{d}\,\partial_k\hat{n}^{d})
\nonumber \\ & & - \frac{m^{3}}{12 \pi(m^{2}-z^{2})^{3}}\int d^{2}x\,
\left[2\vec{\nabla}^{2}(\epsilon^{ij}\, \epsilon^{abc}\, \hat{n}^{a}\, \partial_i \hat{n}^{b}\,
\partial_j \hat{n}^{c})  +3\,\partial_i\left(\epsilon^{ij}\, \epsilon^{abc}\,
(\vec{\nabla}^{2}\hat{n}^{a})\,\partial_j\hat{n}^{b}\,\hat{n}^{c}\right)\right]
\label{derivative4}
\end{eqnarray}
There are several important comments to be made about this fourth order 
contribution. First, the second term contains total spatial 
derivatives, which vanish after doing the $\vec{x}$ integral. Second,
the first term is nontopological and cannot be expressed as a total
derivative. This term gives a nonzero contribution to the spatial
integral, and furthermore this contribution (unlike the winding number
integral in (\ref{fn2})) depends explicitly on the length scale of the
background field
$\hat{n}(\vec{x})$. Third, to see how this is compatible with the fact that the zero
$T$ fermion number is topological, we observe that after doing the $z$
integral in (\ref{contour}), we obtain
\begin{equation}
\langle N\rangle _T^{(4)}=\left[\frac{m^3 T}{12 \pi} \sum_{n=-\infty}^\infty 
\frac{m^2 -5((2n+1)\pi T)^2}{[m^2 +((2n+1)\pi T)^2)]^4}\right]\int d^{2}x\,(\epsilon^{ij}\,
\epsilon^{abc}\, \hat{n}^{a}\, \partial_i\hat{n}^{b}\, \partial_j \hat{n}^{c})(\partial_k
\hat{n}^{d}\, \partial_k \hat{n}^{d})
\label{nt4}
\end{equation}
The prefactor is T dependent, as was the prefactor in the 2-derivative
contribution in (\ref{nt2}).  However, using (\ref{identity}),  as
$T\to 0$, the prefactor of the fourth order term (\ref{nt4}) behaves as
\begin{eqnarray}
 \frac{m^3 T}{12\pi} \sum_{n=-\infty}^\infty \frac{m^2-5((2n+1)\pi
T)^2}{[m^2+((2n+1)\pi T)^2]^4} & \sim &  \frac{m^3}{12
\pi}\left[\left(\frac{15}{16 m^5}-\frac{1}{8 m^2
T^3}e^{-\frac{m}{T}}+\dots\right)-\left(\frac{15}{16 m^5}-\frac{5}{8
m^3 T^2}e^{-\frac{m}{T}}+\dots\right) \right] \nonumber \\ & \sim  &
-\left(\frac{m}{96\pi
T^3}\right)\,e^{-\frac{m}{T}}\,\left(1-5\frac{T}{m}+\dots\right)
\label{4der}
\end{eqnarray}
so that it vanishes (exponentially) as $T\to 0$.  Thus, at $T=0$ the 
nontopological contribution to the  fermion number vanishes, at this
order in the derivative expansion.  Note that in (\ref{2der})
the leading constant term 1 survives the $T\to 0$ limit, while  in
(\ref{4der}) the leading constant terms cancel, leaving a
function that vanishes as $T\to 0$. 

Already at this next-to-leading order of the derivative expansion, we
have established in (\ref{nt4}) that the induced fermion number acquires
a nontopological temperature dependent contribution at finite T. We now
consider higher orders in the derivative expansion. The next term beyond
(\ref{derivative4}) has 6 derivatives and receives contributions from
each of the first five terms in the expansion (\ref{expansion}). While
it is systematic to write down all these 6-derivative terms, it
becomes a lengthy expression due to the many ways of contracting the
spacetime and SU(2) indices. This proliferation of terms
rapidly becomes worse as we go to higher and higher orders in the
derivative expansion. To proceed, we consider instead the low temperature
limit. In this case we find that we can evaluate the leading contribution
to all orders of the derivative expansion, and resum this leading low
temperature correction to the zero temperature answer (\ref{nzero}).
This follows the procedure used in \cite{ad} for a $1+1$ dimensional
sigma model, and is based on simple dimensional analysis.  From the
general expansion (\ref{expansion}) we see that terms either involve:
(a) an even number of $ V$'s and one interaction term $I$, or (b) an odd
number of $V$'s and one kinetic term $K$. Consider a term of form (a) with
$\nu$ vertex insertions of $V$. Since $V$ has already a derivative of
$\hat{n}(\vec{x})$, the total number of derivatives, $d$, of
$\hat{n}(\vec{x})$ in this term satisfies $2\le \nu
\le d$. This term also has
$\nu +1$ propagators $\triangle$. Thus, by dimensional reasoning, this term must
behave as
\begin{equation}
\textrm{Tr}((\triangle V)^{\nu}\triangle I)^{(d)}\sim 
\frac{m^{\nu +1}}{(m^{2}-z^{2})^{(\nu +d)/2}} \mathcal{A}_{d}^{\nu
+1}[\hat{n}(\vec{x})]
\label{dimensiona}
\end{equation}
where $\mathcal{A}_{d}^{\nu +1}[\hat{n}(\vec{x})]$ is some functional of 
$\hat{n}(\vec{x})$ with d derivatives and $(\nu +1)$ factors of 
$\hat{n}(\vec{x})$.
On the other hand,  a term of the form (b) behaves as 
\begin{equation}
\textrm{Tr}((\triangle V)^{\nu}\triangle K)^{(d)}\sim 
\frac{m^{\nu}}{(m^{2}-z^{2})^{(\nu +d -1)/2}} \mathcal{A}_{d}^{\nu }[\hat{n}(\vec{x})]
\label{dimensionb}
\end{equation}
We now consider the $T$ dependence of the prefactors of these types of 
contributions. First, for $d\geq 4$ the constant terms in the low $T$
($T\ll m$) expansion  must all cancel at any given order $d$ of the
derivative expansion, since the final answer must be just the
2-derivative topological term (\ref{nzero}) at $ T=0$. This is a very
stringent test of the derivative expansion at finite temperature. We have
explicitly checked that this is indeed satisfied for $d=4$ and $d=6$. The
leading corrections all have the same exponential factor, with a prefactor
coefficient that depends on the power of the propagator. From
(\ref{identity}) we see that the dominant prefactor at low temperature
occurs when $p$ (the power to which the propagator is raised) is as large
as possible. This means that for the terms in (\ref{dimensiona}) and
(\ref{dimensionb}) we must take $\nu =d$. This means that each derivative
just comes from one of the insertions of $V$. This fact dramatically simplifies the derivative
expansion (in this low $T$ limit), because the propagator factors can
simply be moved out of the trace without generating further derivatives
(since these would be subleading). We also learn that the terms
of the form (\ref{dimensiona}) dominate over those of the form
(\ref{dimensionb}). These simplifications permit us to go to any order of
the derivative expansion.

To compute the actual form of the derivtive expansion contribution in this
low temperature limit, we  have to calculate
$\mathcal{A}_{d}^{\nu +1}[\hat{n}(\vec{x})]$ in (\ref{dimensiona}), which
involves doing the (matrix) traces over the Dirac and isospin matrices:
\begin{equation}
\textrm{tr}\left[I V^{d}\right]= \textrm{tr}\left[m\,
\sigma^{3}\otimes \left(\hat{n}(\vec{x})\cdot 
\vec{\tau}\right)\left(m^2\,  \partial_{i}\hat{n}^a
\,\partial_{j}\hat{n}^b\, \sigma^{i}\, \sigma^{j} \otimes \tau^{a}\,
\tau^{b}\right)^{\frac{d}{2}}\right]
\end{equation}
where $(\sigma^{3}, -i\vec{\sigma})$ and $\vec{\tau}$ are the Pauli matrices in 
the Dirac and Isospin spaces respectively. Using the fact that $i$ and $ j$ can only
take values $1$ or $2$, and that $\sigma^{i} \sigma^{j}=\delta^{ij}
+i\,\epsilon^{ijk}\,\sigma^{k}$, the above trace becomes
\begin{equation}
m^{d+1}\textrm{ tr}\left[\sigma^{3} \otimes
(\hat{n}(\vec{x})\cdot\vec{\tau})\left(\partial_{i}\hat{n}^{a}\,
\partial_{i}\hat{n}^{a}-\epsilon^{ij}\,\epsilon^{abc}\, \partial_{i}\hat{n}^a\,
\partial_{j}\hat{n}^b\, \sigma^{3} \otimes \tau^{c}\right)^{\frac{d}{2}}\right]
\label{caltrace}
\end{equation}
Using the binomial expansion for expanding the above power and noting
that:  (i) $\textrm{Tr}(\sigma^{3})^{m+1}=2$ when m is odd and zero
otherwise, and (ii)
$(\vec{\tau}\cdot \vec{a})(\vec{\tau}\cdot \vec{b})=\vec{a}\cdot \vec{b}+i
(\vec{a}\times\vec{b})\cdot\vec{\tau}$, the expression (\ref{caltrace})
becomes
\begin{equation}
-4 m^{d+1} \, \, \frac{q(\vec{x})}{|\vec{v}(\vec{x})|} \sum_{m=1, 3, 5, . . }^{d/2} 
{\frac{d}{2} \choose m}(\partial_{i}\hat{n}^{a}\,
\partial_{i}\hat{n}^{a})^{(\frac{d}{2} -m)} |\vec{v}(\vec{x})|^{m}
\end{equation}
where we have defined the following combinations
\begin{equation}
\vec{v}(\vec{x})\equiv \epsilon^{i j}\,\partial_{i} \hat{n} \times
\partial_{j} 
\hat{n}\quad  , \quad q(\vec{x})\equiv\epsilon^{i j}\, \epsilon^{abc}\,
\hat{n}^{a}\,
\partial_{i}\hat{n}^{b}\, \partial_{j} \hat{n}^{c}
\end{equation}
Finally, performing the functional trace over the propagators we find 
\begin{equation}
\textrm{Tr}((\triangle V)^{\nu}\triangle I)^{(d)}\sim -\frac{m^{d+1}}{2
\pi d  (m^2 -z^2)^{d}}\int d^{2}x\,
\frac{q(\vec{x})}{|\vec{v}(\vec{x})|}\left[\left(\partial_{i}\hat{n}^{a}\,
\partial_{i}\hat{n}^{a} +|\vec{v}(\vec{x})|\right)^{d/2} -\left(\partial_{i}\hat{n}^{a}\,
\partial_{i}\hat{n}^{a} -|\vec{v}(\vec{x})|\right)^{d/2}\right]
\label{dt}
\end{equation}
So, in the low $T$ limit ($T\ll m$), the derivative expansion becomes
very compact. The leading low temperature contribution, with $d$ spatial
derivatives, to the induced fermion number is then found by summing over
the Matsubara poles of the prefactor (\ref{dt}), and then using the low
temperature expansion in (\ref{identity}). This gives the following
\begin{equation}
\langle N\rangle_T^{(d)}\sim -
\frac{mT}{\pi} \frac{1}{d! (2T)^d}\,e^{-\frac{m}{T}}\,\int d^2 x\,
\frac{q(\vec{x})}{|\vec{v}(\vec{x})|}\left[\left(\partial_{i} \hat{n}^a \,\partial_{i} \hat{n}^a
+|\vec{v}(\vec{x})|\right)^{\frac{d}{2}}-\left(\partial_{i}\hat{n}^a \, \partial_{i}\hat{n}^a
-|\vec{v}(\vec{x})|\right)^{\frac{d}{2}}\right]
\label{leading}
\end{equation}
Note that at any order $d$ of the derivative expansion, the prefactor term
in (\ref{dt}) leads to a temperature depndent prefactor in the induced
fermion number that vanishes as $T\to 0$. Furthermore, the leading low
$T$ terms in (\ref{leading}) have a simple form that can be resummed to
all orders in the derivative expansion. Combining with the 2-derivative
term in (\ref{nt2}), we obtain the following all-orders result for the low
temperature correction to the induced fermion number in the $(2+1)$
dimensional  nonlinear
$\sigma$ model:
\begin{equation}
\langle N\rangle _{T} \sim \langle N\rangle _{0} -
\frac{m T}{\pi} e^{-\frac{m}{T}}\int d^{2}x \, \frac{q(\vec{x})}{|\vec{v}(\vec{x})|}\left[
\cosh\left(\frac{\sqrt{\partial_{i}\hat{n}^a\, \partial_{i}\hat{n}^a +|\vec{v}(\vec{x})|}}{2
T}\right) -\cosh \left(\frac{\sqrt{\partial_{i}\hat{n}^a\, \partial_{i}\hat{n}^a
-|\vec{v}(\vec{x})|}}{2 T}\right)\right]
\label{exactresult}
\end{equation}
This is the main result of this paper. It shows explicitly that the 
topological zero temperature fermion number, given by the winding number $\langle
N\rangle _{0}$ in (\ref{nzero}) acquires a $T$ dependent correction which
is nontopological. This temperature dependent correction cannot be
expressed in terms of the winding number, and is sensitive to the
detailed shape of the background field $\hat{n}(\vec{x})$, not just its
asymptotic values. Also note that the $T\to 0$ limit is well defined
in the derivative expansion regime where (\ref{deriv}) is satisfied. This
resummed expression (\ref{exactresult}) is analogous to a similar
all-orders resummed result found in \cite{ad} for the $(1+1)$ dimensional
sigma model background. We find it remarkable that this all-orders
resummation can also be done for the $(2+1)$ dimensional sigma model, as
it is a significantly more complicated model.

Thus far, the only thing we have assumed about
$\hat{n}(\vec{x})$ is that it takes values in $ S^{2}$,  see
(\ref{constraint}), and that it has a ``gentle'' spatial profile, see
(\ref{deriv}). We now consider two specific explicit forms for the
background $\hat{n}(\vec{x})$. 

First, suppose $\hat{n}(\vec{x})$ has the form of a 2-dimensional
$\mathcal{CP}^{1}$ instanton, in which case the winding number $\langle
N\rangle_0$ in (\ref{nzero}) is just the instanton number, as is well
known for the zero temperature system \cite{raj}. 
The instanton field $\hat{n}(\vec{x})$ satisfies the
first order instanton equation
\begin{equation}
\partial_{i} \hat{n}^{a}=\pm \epsilon_{ij}\, \epsilon^{abc}\, \hat{n}^{b}\, 
\partial_{j} \hat{n}^{c}
\label{condition}
\end{equation}
where the $\pm$ signs refer to the anti-instanton/instanton cases
respectively (one can easily check this by substituting (\ref{condition})
in the expression (\ref{nzero}) to find the sign of  $\langle N \rangle
_0$). Now, in general we have that $|\vec{v}(\vec{x})|^2 =2(\partial_{i}
\hat{n}^{a}\,\partial_{i}\hat{n}^{a})^{2} -2(\partial_{i}\hat{n}\cdot  
\partial_{j}\hat{n})^{2}$. But for an instanton background $\hat{n}$
satisfying (\ref{condition}), it follows that
$(\partial_{i}\hat{n}\cdot\partial_{j}\hat{n})^{2}=\frac{1}{2}(\partial_{i}\hat{n}
\cdot \partial_{i}\hat{n})^{2}$. Therefore,
$|\vec{v}(\vec{x})|=(\partial_{i}\hat{n}\cdot \partial_{i}
\hat{n})$, and $q(\vec{x})=\mp (\partial_{i}\hat{n} \cdot  \partial_{i}
\hat{n})$. Then the low $T$ expansion (\ref{exactresult}) simplifies
somewhat:
\begin{equation}
\langle N\rangle _{T} \sim \langle N\rangle _{0} \pm 
\frac{m T}{\pi} e^{-\frac{m}{T}} \int d^{2}x\, \left[\cosh\left(\frac{\sqrt{2 \partial_{i}\hat{n}\cdot
\partial_{i} \hat{n}}}{2 T}\right)-1\right]
\label{instantonfn}
\end{equation}

To be even more explicit, we choose $\hat{n}$ to be a 
 $k$-instanton solution, in which case $\langle N \rangle _0=k$. The
explicit $\mathcal{CP}^{1}$ instantons are most easily expressed in terms
of the following stereographically projected fields $\omega^1$ and
$\omega^2$:
\begin{equation}
 \omega^{1}=\frac{2 n^{1}}{(1-n^{3})} \quad , \quad \omega^{2}=
\frac{2 n^{2}}{(1-n^{3})}
\end{equation}
Defining $\omega =\omega ^{1} +i \omega ^{2}$, the
original sigma model field $\hat{n}(\vec{x})$ is
\begin{equation}
\hat{n}(\vec{x})=\frac{1}{\frac{|\omega |^{2}}{4} +1}\left(\begin{array}{ccc} 
\textrm{Re}
\,\omega \\ \textrm{Im}\,\omega \\ \frac{|\omega|^{2}}{4} -1 \end{array}
\right)
\label{instfield}
\end{equation}
As is well known, the instanton equation (\ref{condition}) then becomes a
simple Cauchy-Riemann condition for $\omega$, so that instantons are
characterized by a meromorphic function of $z=x^1+i x^2$. For example,
\begin{equation}
\omega(z)=\left[\frac{(z-z_{0})}{\lambda }\right]^{k}
\end{equation}
represents $k$ instantons at the location $z_0$, each with a common
length scale $\lambda$. For such an instanton,
\begin{equation}
\partial _{i}\hat{n}\cdot \partial _{i}\hat{n}=\left( \frac{2 k^{2}}{\lambda ^{2k}}\right) 
|z -z_{0}|^{2k-2} \frac{1}{\left[1+\frac{1}{4 \lambda^{2k}}
|z-z_{0}|^{2 k}\right]^{2}}
\end{equation}
The spatial integrals in (\ref{instantonfn}) can now be done, and one
finds
\begin{equation}
\langle N\rangle _{T}\sim k \pm \frac{4k m}{T} e^{-\frac{m}{T}} 
\sum_{n=0}^\infty \frac{1}{(2 n+2)!}\left(\frac{2 k}{2^{\frac{1}{k}} \lambda T}\right)^{2 n}
\beta\left(n+1+\frac{n}{k}, n+1-\frac{n}{k}\right)
\label{spaceintegral}
\end{equation}
where $\beta(m, n)=\frac{\Gamma(m) \Gamma(n)}{\Gamma(m+n)}$ is the
Euler beta function. The first term, $k$, in (\ref{spaceintegral}) is the
zero temperature fermion number $\langle N \rangle _{0}$, which is simply
the instanton number $k$, and which is manifestly independent of the
instanton scale $\lambda$. On the other hand, the finite $T$ correction
in (\ref{spaceintegral}) is manifestly dependent on the instanton scale
$\lambda$, reflecting its nontopological nature. It is interesting to
note that this correction term is a convergent series in $\frac{1}{\lambda
T}$, for any instanton number $k$. For a single instanton $(k=1)$, it
simplifies further to
\begin{eqnarray}
\langle N\rangle _{T}& \sim & 1 - \frac{4 m}{T} e^{-\frac{m}{T}}\sum_{n=0}^\infty 
\frac{1}{(2 n +1)(2 n+2)!} \left(\frac{1}{\lambda T}\right)^{2 n}\nonumber \\
 {} &=&1 - (4 m \lambda)e^{-\frac{m}{T}}\int_{0}^{\frac{1}{\lambda T}} 
\frac{\cosh y -1}{y^2} dy 
\label{fninst}
\end{eqnarray}

The second physically interesting form for the sigma model background
field $\hat{n}$ is the ``baby Skyrmion'' case
\begin{equation}
\hat{n}(r, \phi)=\left(\begin{array}{ccc} \cos \phi \sin \theta (r) \\ \sin \phi 
\sin \theta (r) \\ \cos \theta (r)
\end{array} \right)
\label{baby}
\end{equation}
Here, $r$ and $\phi$ are the 2-dimensional polar coordinates and $\theta
(r)$ is  some radial profile function. The ``baby Skyrmion'' is the $2$
dimensional analog of the $3$ dimensional Skyrme model for baryons, with a
``hedgehog'' ansatz for the meson fields \cite{weigel}. For such an
ansatz, the winding number density is
\begin{equation}
q(\vec{x})=\epsilon ^{abc}\,\epsilon ^{ij}\, \hat{n}^{a}\, \partial_{i} \hat{n}^{b}\, \partial_{j} 
\hat{n}^{c}=\frac{2}{r} \sin \theta (r)\, \theta^{\prime}(r)
\end{equation}
so that the zero temperature induced fermion number (\ref{nzero}) is
\begin{equation}
\langle N\rangle _0 =\frac{1}{8 \pi}\left[4\pi \int_{0}^{\infty}dr \frac{d\theta }{dr} 
\sin \theta (r)\right]=\frac{1}{2}\left[\cos \theta (0)-\cos \theta (\infty )\right]
\end{equation}
This shows clearly that $\langle N\rangle _0 $ is topological : it only 
depends on the asymptotic values of the profile function $\theta (r)$, not on its
specific spatial profile or scale. The low temperature correction to the fermion
number in this case is obtained from (\ref{exactresult}) by noting that 
for the ansatz (\ref{baby}),
\begin{eqnarray}
\partial_{i} \hat{n}^{a} \partial_{i} \hat{n}^{a}
&=&(\theta^{\prime}(r))^{2}  +\left(\frac{\sin \theta(r)}{r}\right)^{2}
\nonumber\\
 (\partial_{i} \hat{n}\cdot
\partial_{j}\hat{n})^{2}&=&(\theta^\prime (r))^{4} +\left(\frac{\sin
\theta (r)}{r}\right)^{4} 
\end{eqnarray}
in which case, 
\begin{equation}
|\vec{v}(r, \phi )|=\frac{2  |\theta^{\prime}(r) \sin \theta (r)|}{r}
\end{equation}
Thus,  the low temperature induced fermion number for the 
``baby Skyrmion'' background  is
\begin{equation}
\langle N \rangle _{T} \sim \langle N \rangle _{0} -
2 m T  e^{-\frac{m}{T}} \int_{0}^{\infty}r\, dr\,  
\textrm{sign}\left(\theta^{\prime}(r) \sin \theta (r)\right)\,  \left[
\cosh\left(\frac{|\theta^{\prime}(r)| +\frac{|\sin \theta (r)|}{r}}{2
T}\right) -
\cosh\left(\frac{|\theta^{\prime}(r)| -\frac{|\sin \theta (r)|}{r}}{2
T}\right)\right]
\label{fnbaby}
\end{equation}
The background field $\theta (r)$ has to satisfy certain boundary
conditions. Requiring that, for a localized soliton solution, the chiral
field as $r \to \infty$ should equal its vacuum value, i.e, $\hat{n}(r,\phi)$ should be independent of the angle $\phi$, fixes
$\theta(\infty)=k \pi$, where $k$ is an integer. At the origin,
finiteness of the soliton energy requires $\theta(0)=n \pi$, where $n$ is
an integer. As an example, consider the
following ansatz for the radial profile function
$\theta(r)$:
\begin{equation}
\theta(r)=2\arctan \left(\frac{r}{2 \lambda}\right)
\label{ansatz}
\end{equation}
Then 
\begin{equation}
\theta^{\prime}(r)=\frac{4 \lambda}{r^{2} +4
\lambda^{2}}=\frac{\sin \theta(r)}{r}
\end{equation} 
Substituting this in (\ref{fnbaby}) we obtain
\begin{equation}
\langle N \rangle _{T} \sim 1 - 
4 m \lambda e^{-\frac{m}{T}} \int_{0}^{\frac{1}{\lambda T}}dy 
\frac{\cosh y -1}{y^{2}}
\end{equation}
which is, incidentally, exactly the finite temperature fermion number
(\ref{fninst}) for the $\mathcal{CP}^1$ single instanton background.

The most important thing to be learned from the baby Skyrmion result
(\ref{fnbaby}) is that while the zero temperature fermion number $\langle
N\rangle_0$ only depends on the asymptotic values of the radial profile
function $\theta(r)$, the finite $T$ correction is sensitive to the
details of the shape of $\theta(r)$. This means, for example, that in a
variational calculation where the asymptotic values of $\theta(r)$ are
fixed (to keep the Skyrme number fixed), but the shape of $\theta(r)$ is
varied, the finite temperature fermion number will vary.

\section{Conclusions}

To conclude, we have computed the finite temperature induced fermion number for
fermions in a static nonlinear $\sigma$-model background in $(2+1)$ dimensions. This
calculation illustrates the splitting of the induced fermion number into a
zero temperature piece, which is topological (here, the winding number of
the sigma model background), and a finite temperature correction which is
nontopological \cite{dr}. The calculation was done in the derivative
expansion limit where the spatial derivatives of the background fields
are assumed to be small compared to the fermion mass scale parameter $m$.
We found that it is possible to resum the derivative expansion to all
orders, in the low temperature limit where $T\ll m$. Such a resummation
was done previously \cite{ad} in the $(1+1)$ dimensional $\sigma$-model
case, but we find it remarkable that it can also be done in the more
complicated $(2+1)$ dimensional case. These results were then applied to two
specific background fields in the $\sigma$-model case : a 2-dimensional
$\mathcal{CP}^{1}$ instanton, and a 2-dimensional ``baby Skyrmion''. In
each case, the finite temperature correction can be computed in the low
temperature limit. For an instanton background the final expression for
the induced fermion number
$\langle N\rangle _{T}$ becomes a simple convergent series in $\frac{1}{\lambda T}$
where $\lambda$ is the length scale of the instanton, thus explicitly illustrating the
nontopological character of the finite temperature corrections in $\langle N\rangle
_{T}$.  

Finally, we make several comments on possible extensions of this work. As has been
emphasized recently at zero temperature in \cite{aw,abanov}, the induced charge (and,
more generally, induced current) is only one part of the induced effective action for
fermions in a $\sigma$-model background. There is, in addition, a Hopf term which is
topological, but which does not contribute to the induced current (since
this is defined through a functional variation of the effective action
with respect to a gauge field). This Hopf term has important implications
for the spin and statistics of solitons \cite{wilczekzee,aw,abanov}. It
would be interesting to investigate these issues at finite temperature,
in the language of finite $T$ effective actions. Another result from our
calculation is the implication it has for the chiral sigma models in
$(3+1)$ dimensions. There, the zero temperature calculation is very similar to the $(2+1)$
dimensional case. At finite temperature we learn from our calculation that there will
be a finite temperature correction and that it will be nontopological -- that is,
while the zero temperature induced fermion number is the Skyrme number, the
temperature dependent correction will be a much more complicated functional of the
Skyrme background. This is already clear from the general argument about finite $T$
fermion number in \cite{dr}, together with the numerical results (see, e.g.,
\cite{ripka}) for the sensitive dependence of the fermion spectrum on the scale of
the Skyrme background. That is, since the single-particle fermion spectrum
is not symmetric and is sensitive to the scale of the background, the
finite temperature correction is also sensitive to this scale, and thus
must be nontopological. From a calculational standpoint, we learn that
while at zero temperature the induced fermion number arises from the
leading nontrivial order of the derivative expansion
\cite{eric,ian}, at finite temperature it is necessary to go beyond this
order. In other words, at finite temperature there are contributions that
are not total derivatives, and these contribute to the integrated charge.
It is a much more difficult problem to investigate higher orders of the
derivative expansion in the $(3+1)$ dimensional Skyrme model. However,
the results of this paper for the $(2+1)$ dimensional model suggest
that it should be possible to explore higher order of the derivative
expansion, at least in the low temperature limit. It would also be of
interest in this context to extend the systematic renormalization approach
developed in \cite{jaffe2,noah}, which is ideally suited to numerical
evaluation of induced charges, to finite temperature.
 
\begin{acknowledgments}
GD and KR thank the U.S. Department of Energy for support through grant
DE-FG02-92ER40716. 00.  JL thanks the Spanish Ministerio de Educaci\'on y
Cultura for support and the UConn Physics Department for hospitality
while this work was done. GD thanks N. Graham, R. L. Jaffe and H. Weigel
for helpful discussions about induced charges.
\end{acknowledgments}


\end{document}